\documentclass[twocolumn,english,aps,pre]{revtex4}
\usepackage[T1]{fontenc}
\usepackage[latin9]{inputenc}
\usepackage{babel}
\usepackage{amsmath}
\usepackage{amssymb}
\usepackage{graphicx}
\usepackage[pdfusetitle,
 bookmarks=true,bookmarksnumbered=true,bookmarksopen=false,
 breaklinks=false,pdfborder={0 0 1},backref=section,colorlinks=false]
 {hyperref}

\makeatletter

\newcommand{\lyxmathsym}[1]{\ifmmode\begingroup\def\b@ld{bold}
  \text{\ifx\math@version\b@ld\bfseries\fi#1}\endgroup\else#1\fi}

\@ifundefined{textcolor}{}
{%
 \definecolor{BLACK}{gray}{0}
 \definecolor{WHITE}{gray}{1}
 \definecolor{RED}{rgb}{1,0,0}
 \definecolor{GREEN}{rgb}{0,1,0}
 \definecolor{BLUE}{rgb}{0,0,1}
 \definecolor{CYAN}{cmyk}{1,0,0,0}
 \definecolor{MAGENTA}{cmyk}{0,1,0,0}
 \definecolor{YELLOW}{cmyk}{0,0,1,0}
}

\makeatother

\begin{document}
\title{Dual thermal pseudocritical features in a spin-1/2 Ising chain with
twin-diamond geometry}
\author{Onofre Rojas}
\affiliation{Department of Physics, Institute of Natural Science, Federal University
of Lavras, Lavras, MG, Brazil}
\begin{abstract}
We study the coupled twin-diamond chain, a decorated one-dimensional
Ising model motivated by the magnetic structure of $\mathrm{Cu}_{2}(\mathrm{TeO}_{3})_{2}\mathrm{Br}_{2}$.
By applying an exact mapping to an effective Ising chain, we obtain
the full thermodynamic description of the system through a compact
transfer-matrix formulation. The ground-state analysis reveals five
distinct phases, including two frustrated sectors with extensive degeneracy.
These frustrated regions give rise to characteristic entropy plateaus
and separate the ordered phases in the zero-temperature diagram. At
low temperatures the model exhibits peculiar sharp yet continuous
variations of entropy, magnetization, and response functions, reflecting
clear signatures of pseudotransition behavior. The coupled twin-diamond
chain thus provides an exactly solvable setting in which competing
local configurations and internal frustration lead to pronounced dual
pseudocritical features in one dimension.
\end{abstract}
\maketitle

\section{Introduction.}

One-dimensional lattice models have recently attracted renewed attention
because certain systems display thermodynamic anomalies that closely
resemble finite-temperature phase transitions \citep{galisova,strecka16,sm-souza16}.
These anomalies, known as pseudotransitions, differ from genuine criticality
even though they reproduce several of its characteristic signatures.
First derivatives of the free energy, such as entropy and magnetization,
exhibit steep but continuous variations near a pseudocritical temperature,
while second derivatives, including the specific heat and magnetic
susceptibility, develop sharp yet finite peaks \citep{sergio18}.
Numerical and analytical studies of decorated Ising and Ising-Heisenberg
chains have further shown that these pseudotransitions can mimic essential
features of critical phenomena: the correlation length increases rapidly
near the pseudocritical temperature \citep{Icarvalho}, and several
response functions follow power-law forms governed by a universal
set of pseudocritical exponents identified in many one-dimensional
decorated models \citep{Rojas2019}.

Peculiar decorated one-dimensional spin models provide a natural framework
for these anomalies because their internal degrees of freedom generate
competing low-energy sectors whose exchange of dominance at a characteristic
temperature produces fully analytic yet remarkably sharp thermodynamic
crossovers. This mechanism has been identified in several exactly
solvable systems, including the Ising-XYZ diamond chain, where quasi-phases
and pseudotransitions were first clearly characterized \citep{sergio18,Icarvalho},
the coupled spin-electron double-tetrahedral chain, which exhibits
frustration-enhanced entropy plateaus \citep{galisova}, and various
Ising-Heisenberg ladders and tubes that display pseudocritical behavior
governed by universal exponents \citep{strecka16,Rojas2019,Derzhko,dmd-potts,Krokhmalskii,Chapman}.
A unified interpretation based on the decoration-iteration mapping
clarified that the anomalous crossover originates from temperature-dependent
effective couplings and effective fields generated after integrating
out the decorated sublattices \citep{Rojas2019}. Furthermore, subsequent
analyses established that the residual entropy at zero-temperature
phase boundaries serves as a reliable predictor for the appearance
of pseudotransitions in decorated chains \citep{bjp20}.

Pseudotransition behavior is not restricted to decorated chains. Several
strictly one-dimensional models without auxiliary spins or internal
bond structures also display sharp thermal crossovers when their parameters
are tuned to generate competing low-energy configurations. A notable
example is the extended Hubbard model at half filling in the atomic
limit, where pronounced thermodynamic anomalies mimic those associated
with quasi-phases \citep{verissimo}. A similar crossover between
antiferromagnetic and charge-ordered configurations was reported in
a minimal spin-pseudospin description of cuprate chains, accompanied
by a strong but finite specific heat peak \citep{katarina}. Related
effects occur in frustrated Potts-type models such as the q-state
Potts-Zimm-Bragg formulation \citep{Panov2021}, in dilute Ising chains
with impurity induced frustration \citep{Yasinskaya}, and in water-like
lattice models incorporating van der Waals interactions and directional
bonding \citep{Braz25}. These examples demonstrate that the essential
ingredients behind pseudotransitions, namely competing sectors with
small energy gaps and enhanced degeneracies, can arise naturally even
in purely one-dimensional systems without explicit decoration.

Diamond chain geometries play a central role in several magnetic compounds,
which explains why diamond chain models have been extensively studied
in the theoretical literature \citep{sm-souza16,strecka-dmd,canova04,Dmd-ch-Strecka22}.
A prominent example is the natural mineral azurite $\mathrm{Cu_{3}(CO_{3})_{2}(OH)_{2}}$\citep{kikuchi03,kikuchi04},
whose magnetic properties motivated detailed analyses of spin 1/2
Ising and Ising-Heisenberg diamond chains. Other compounds, such as
the mixed valent iron diamond chain with single ion anisotropy \citep{Sorolla},
further support the idea that the diamond motif is an efficient source
of frustration and anomalous thermodynamic behavior in one dimensional
systems, among other similar structures\citep{Derzhko,zad,Chakhmakhchyan}.
For these reasons, diamond chains remain key reference models for
exploring the origin of entropy plateaus, degeneracy driven anomalies,
and pseudocritical behavior in one dimension.

Beyond simple diamond chains, an increasing number of materials display
coupled diamond like building blocks, often arranged in layers or
extended structural networks. A representative case is the oxohalide
$\mathrm{Cu}_{2}(\mathrm{TeO}_{3})_{2}\mathrm{Br}_{2}$, where $\mathrm{Cu}^{2+}$
ions form diamond shaped units that are connected along a crystallographic
direction \citep{Uematsu}. Similar motifs appear in more complex
families such as the $\mathrm{CuO\lyxmathsym{\textendash}CuCl_{2}\lyxmathsym{\textendash}SeO_{2}}$
series \citep{Kakarla}, as well as in the oxohalides $\mathrm{Cu}_{3}(\mathrm{SeO}_{3})_{2}\mathrm{Cl}_{2}$
and $\mathrm{Cu}_{3}(\mathrm{TeO}_{3})_{2}\mathrm{Br}_{2}$ \citep{Rocquefelte},
and the insulating spin-gap compound $\mathrm{CdCu_{2}(SeO_{3})_{2}Cl_{2}}$
\citep{Murtazoev}. Such structural features strongly motivate the
study of theoretical models that incorporate more than one diamond
unit per cell. The coupled twin diamond chain examined in the present
work represents the minimal extension capable of capturing this additional
connectivity and allows us to investigate how the interaction between
two diamond units within each cell can produce multiple frustrated
sectors and distinct pseudocritical features in a single exactly solvable
framework.

In the present work, the coupled twin-diamond chain is therefore introduced
as a minimal theoretical model inspired by such diamond-based magnetic
structures\citep{Uematsu}, rather than as a quantitative description
of any specific compound. The central novelty of the coupled twin-diamond
chain lies in the fact that its minimal unit cell supports more than
one frustrated low-energy manifold with different extensive degeneracies.
As a result, the system exhibits two well-separated pseudocritical
temperature scales associated with successive entropy-driven crossovers
between competing configurations. This provides a clean exactly solvable
example in which dual pseudotransition features arise within a single
one-dimensional decorated Ising geometry, rather than from tuning
parameters between distinct models. We further show that these features
remain robust even when the two spin sublattices respond differently
to an external magnetic field.

This work is organized as follows. Sect. II introduces the Hamiltonian,
outlines the transfer-matrix solution, and presents the zero-temperature
phase diagram together with the corresponding pseudotransition features.
Sect. III examines the thermodynamic behavior in the anomalous region
where these pseudotransitions emerge. Finally, Sec. IV summarizes
the main conclusions.

\section{Coupled twin-diamond chain model}

\begin{figure}[h]
\begin{centering}
\includegraphics[scale=0.9]{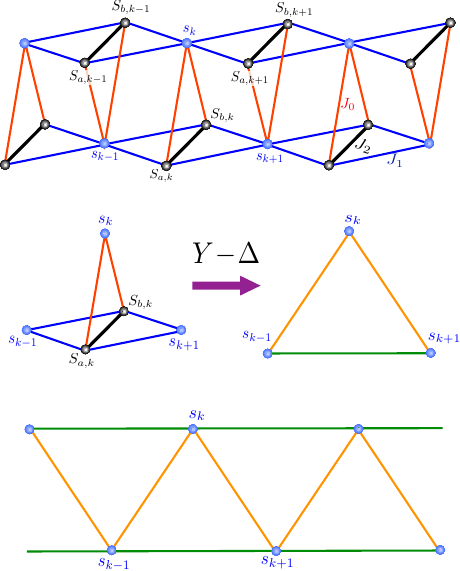}
\par\end{centering}
\caption{\label{fig:CTDC}(Top) Schematic representation of the spin-1/2 CTDC.
Black spheres denote dimer spins ($S_{a,k}$ and $S_{b,k}$), while
blue spheres represent nodal Ising spins. Blue (red) lines indicate
the couplings between dimer and nodal spins, associated with the exchange
interactions $J_{1}$ ($J_{0}$), respectively, whereas the black
line corresponds to the intra-dimer exchange interaction ($J_{2}$).
(Middle) Illustration of the star-triangle transformation. (Bottom)
Resulting effective spin chain with nearest- and next-nearest-neighbor
interactions.}
\end{figure}

In this section we consider a coupled twin-diamond chain (CTDC) inspired
by the compound ${\rm Cu}_{2}({\rm TeO}_{3})_{2}{\rm Br}_{2}$ whose
magnetic structure consists of spin-1/2 ${\rm Cu}^{2+}$ ions arranged
in diamond-like units\citep{Uematsu}. Here we assume the simplest
case that the magnetic interactions are purely of the Ising type.
A schematic representation of the lattice geometry is shown in Fig.
\ref{fig:CTDC}. Blue spheres denote the nodal spins, whereas black
spheres indicate the two spins forming each internal dimer. 

Each unit cell of the CTDC contains a nodal Ising spin $s_{k}$ and
a pair of Ising spins $S_{a,k}$ and $S_{b,k}$. The local fields
acting on the two types of spins may differ because the corresponding
gyromagnetic factor for nodal spins $g_{0}$ and dimer spins $g_{1}$.
The resulting Zeeman terms are $h_{0}=\mu_{\mathrm{B}}g_{0}B$ and
$h_{1}=\mu_{\mathrm{B}}g_{1}B$, which act on the nodal and dimer
spins, respectively.

\subsection{Hamiltonian and transfer matrix}

The Hamiltonian of the coupled twin-diamond Ising chain reads

\begin{alignat}{1}
\mathcal{H}= & \sum\limits_{k=1}^{N}\left\{ \left[J_{0}s_{k}+J_{1}\left(s_{k-1}+s_{k+1}\right)\right]\left(S_{a,k}+S_{b,k}\right)\right.\nonumber \\
 & \left.+J_{2}S_{a,k}S_{b,k}-h_{1}\left(S_{a,k}+S_{b,k}\right)-h_{0}s_{k}\right\} .\label{eq:Ham1}
\end{alignat}

Here $J_{2}$ is the intra-dimer coupling, $J_{0}$ couples the nodal
spin $s_{k}$ to the adjacent dimer, and $J_{1}$ couples the dimer
to the neighboring nodal spins $s_{k-1}$ and $s_{k+1}$ along the
chain. Due to $g_{0}\neq g_{1}$ in general, the two sublattices respond
differently to the same external magnetic field $B$. The Hamiltonian
\eqref{eq:Ham1} is invariant under the exchange $S_{a,k}\leftrightarrow S_{b,k}$
and is translationally invariant along the chain. 

At this stage, the internal dimer spins $S_{a,k}$ and $S_{b,k}$
can be eliminated by means of a local star-triangle transformation
\citep{dec-trns,ddec-trns,jozef-dctrns}, as schematically illustrated
in Fig. \ref{fig:CTDC} (middle). For fixed values of the neighboring
nodal spins, the Boltzmann weight of each decorated cell is obtained
by explicitly summing over the dimer degrees of freedom. The repeated
application of this mapping along the chain yields an effective one-dimensional
Ising model for the nodal spins, with temperature-dependent nearest-
and next-nearest-neighbor interactions and an effective field.

The resulting effective model can be treated exactly by the transfer-matrix
method. Because the mapping generates next-nearest-neighbor couplings,
the transfer matrix $\mathbf{V}$ acts on pairs of adjacent nodal
spins and therefore has dimension $4\times4$. Analogous to the transfer
matrix of the extended Ising chain \citep{stephenson}, $\mathbf{V}$
is written in the pair basis $\{|++\rangle,|+-\rangle,|-+\rangle,|--\rangle\}$
as 

\begin{equation}
\mathbf{V}=\left(\begin{array}{cccc}
\mathfrak{r} & \mathfrak{m} & 0 & 0\\
0 & 0 & \mathfrak{p} & \mathfrak{n}\\
\mathfrak{m} & \mathfrak{q} & 0 & 0\\
0 & 0 & \mathfrak{n} & \mathfrak{t}
\end{array}\right),\label{eq:V}
\end{equation}
where the basis states are labeled by the nodal-spin pairs ($s_{k},s_{k+1}$).
The matrix elements $\mathfrak{r}$, $\mathfrak{m}$, $\mathfrak{p}$,
$\mathfrak{n}$, $\mathfrak{q}$, $\mathfrak{t}$ incorporate the
local Boltzmann weights for the four possible configurations of neighboring
nodal spins and the internal dimer
\begin{alignat}{1}
\mathfrak{r}= & 2{\rm e}^{-\frac{J_{2}+2h_{0}}{4k_{{\rm B}}T}}\cosh\left(\tfrac{J_{0}+2J_{1}+h_{1}}{k_{{\rm B}}T}\right)+2{\rm e}^{\frac{J_{2}-2h_{0}}{4k_{{\rm B}}T}},\\
\mathfrak{m}= & 2{\rm e}^{-\frac{3J_{2}+2h_{0}}{12k_{{\rm B}}T}}\cosh\left(\tfrac{J_{0}+h_{1}}{k_{{\rm B}}T}\right)+2{\rm e}^{\frac{3J_{2}-2h_{0}}{12k_{{\rm B}}T}},\\
\mathfrak{p}= & 2{\rm e}^{-\frac{3J_{2}+2h_{0}}{12k_{{\rm B}}T}}\cosh\left(\tfrac{2J_{1}-J_{0}+h_{1}}{T}\right)+2{\rm e}^{\frac{3J_{2}-2h_{0}}{12k_{{\rm B}}T}},\\
\mathfrak{n}= & 2{\rm e}^{-\frac{3J_{2}-2h_{0}}{12k_{{\rm B}}T}}\cosh\left(\tfrac{J_{0}-h_{1}}{k_{{\rm B}}T}\right)+2{\rm e}^{\frac{3J_{2}+2h_{0}}{12k_{{\rm B}}T}},\\
\mathfrak{q}= & 2{\rm e}^{-\frac{3J_{2}-2h_{0}}{12k_{{\rm B}}T}}\cosh\left(\tfrac{2J_{1}-J_{0}-h_{1}}{k_{{\rm B}}T}\right)+2{\rm e}^{\frac{3J_{2}+2h_{0}}{12k_{{\rm B}}T}},\\
\mathfrak{t}= & 2{\rm e}^{-\frac{J_{2}-2h_{0}}{4k_{{\rm B}}T}}\cosh\left(\tfrac{2J_{1}+J_{0}-h_{1}}{k_{{\rm B}}T}\right)+2{\rm e}^{\frac{J_{2}+2h_{0}}{4k_{{\rm B}}T}},
\end{alignat}

The eigenvalues $\lambda_{i}(i=1,\dots,4)$ follow from the secular
equation $\det(V-\lambda I)=0$, which yields a quartic polynomial
\begin{equation}
\lambda^{4}+\mathfrak{c}_{3}\lambda^{3}+\mathfrak{c}_{2}\lambda^{2}+\mathfrak{c}_{1}\lambda+\mathfrak{c}_{0}=0,
\end{equation}
where the coefficients are given by 
\begin{alignat}{1}
\mathfrak{c}_{0}= & -\left(\mathfrak{q}\mathfrak{r}-\mathfrak{m}^{2}\right)\left(\mathfrak{p}\mathfrak{t}-\mathfrak{n}^{2}\right)<0,\\
\mathfrak{c}_{1}= & \left(\mathfrak{q}\mathfrak{r}-\mathfrak{m}^{2}\right)\mathfrak{p}+\left(\mathfrak{p}\mathfrak{t}-\mathfrak{n}^{2}\right)\mathfrak{q}>0,\\
\mathfrak{c}_{2}= & \mathfrak{r}\mathfrak{t}-\mathfrak{p}\mathfrak{q},\\
\mathfrak{c}_{3}= & -(\mathfrak{r}+\mathfrak{t})<0.
\end{alignat}

To simplify the quartic polynomial, we introduce the shift $\lambda=u+\frac{\mathfrak{r}+\mathfrak{t}}{4}$,
which eliminates the cubic term and leads to the depressed quartic
equation
\begin{equation}
u^{4}+a_{2}u^{2}+a_{1}u+a_{0}=0,\label{eq:deps-4tic}
\end{equation}
with 
\begin{alignat*}{1}
a_{0}= & -3\frac{\mathfrak{c}_{3}^{4}}{4^{4}}+\frac{\mathfrak{c}_{2}\mathfrak{c}_{3}^{2}}{4^{2}}-\frac{\mathfrak{c}_{1}\mathfrak{c}_{3}}{4}+\mathfrak{c}_{0},\\
a_{1}= & \frac{\mathfrak{c}_{3}^{3}}{2^{3}}-\frac{\mathfrak{c}_{2}\mathfrak{c}_{3}}{2}+\mathfrak{c}_{1},\\
a_{2}= & -3\frac{\mathfrak{c}_{3}^{2}}{8}+\mathfrak{c}_{2}<0.
\end{alignat*}
The four roots of the quartic polynomial can then be written in analytic
form as
\begin{alignat}{1}
\lambda_{1}= & \frac{\mathfrak{r}+\mathfrak{t}}{4}+\frac{\sqrt{2y-a_{2}}+\sqrt{-2y-a_{2}-\frac{2a_{1}}{\sqrt{2y-a_{2}}}}}{2},\label{eq:L1}\\
\lambda_{2}= & \frac{\mathfrak{r}+\mathfrak{t}}{4}+\frac{\sqrt{2y-a_{2}}-\sqrt{-2y-a_{2}-\frac{2a_{1}}{\sqrt{2y-a_{2}}}}}{2},\label{eq:L2}\\
\lambda_{3}= & \frac{\mathfrak{r}+\mathfrak{t}}{4}-\frac{\sqrt{2y-a_{2}}-\sqrt{-2y-a_{2}+\frac{2a_{1}}{\sqrt{2y-a_{2}}}}}{2},\\
\lambda_{4}= & \frac{\mathfrak{r}+\mathfrak{t}}{4}-\frac{\sqrt{2y-a_{2}}+\sqrt{-2y-a_{2}+\frac{2a_{1}}{\sqrt{2y-a_{2}}}}}{2},
\end{alignat}
where the $y$ is choosing for conveniency as

\begin{alignat}{1}
y= & 2\sqrt{Q}\cos\left(\tfrac{1}{3}\arccos(\tfrac{R}{\sqrt{Q^{3}}})\right)+\frac{a_{2}}{6},\label{eq:y-sol}
\end{alignat}
with 
\begin{alignat}{1}
Q= & \frac{a_{2}^{2}}{36}+\frac{a_{0}}{3},\\
R= & \frac{a_{2}^{3}}{216}-\frac{a_{2}a_{0}}{6}+\frac{a_{1}^{2}}{16}.
\end{alignat}

It is worth noting that the largest eigenvalue $\lambda_{1}$ is always
real and strictly positive. On the branch associated with the dominant
Perron\citep{cuesta} eigenvalue, the second eigenvalue $\lambda_{2}$
is also real and satisfies $\lambda_{1}>\lambda_{2}$, as demonstrated
in Appendix \ref{sec:appx-A}. The remaining eigenvalues $\lambda_{3}$
and $\lambda_{4}$ may be either real or form a complex-conjugate
pair, depending on the Hamiltonian parameters. Moreover, no general
ordering can be inferred for the magnitudes $|\lambda_{2}|$, |$\lambda_{3}|$,
and $|\lambda_{4}|$, which may vary across parameter regimes (see
Appendix \ref{sec:appx-A}).

Among these, the largest eigenvalue can be identified as $\lambda_{1}$.
Thus in the thermodynamic limit, the free energy per unit cell is
$f(T)=-k_{{\rm B}}T\ln\lambda_{1}$, from which the entropy $\mathcal{S}=-\partial f/\partial T$,
the specific heat $C=T\,\partial\mathcal{S}/\partial T$, and all
other thermodynamic quantities follow in a standard way.

\section{Zero-temperature phases and phase diagram}

\begin{figure}
\includegraphics[scale=0.7]{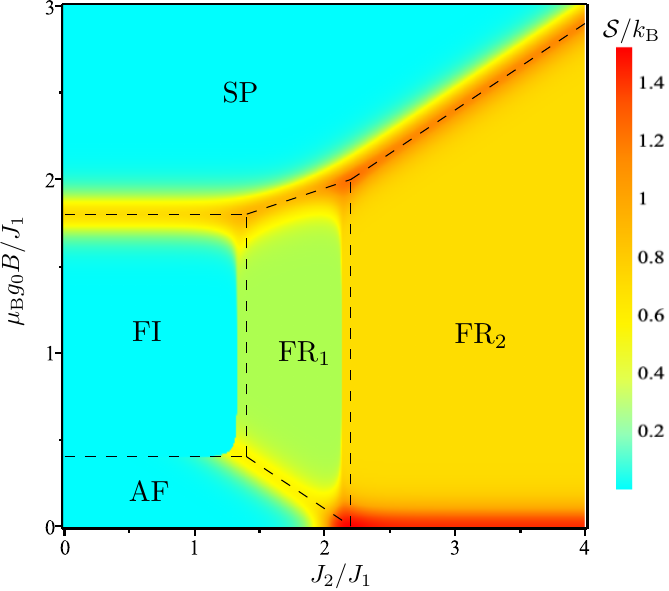}\caption{\label{fig:Ph-dgm1} Zero-temperature phase diagram and low-temperature
entropy density for the CTDC model. The phase diagram is shown in
the ($J_{2}/J_{1},B/J_{1}$) plane for fixed $J_{0}=-0.2$, and equal
gyromagnetic factors $g_{0}=g_{1}$, so that the Zeeman fields satisfy
$h_{0}=h_{1}=g_{0}\mu_{\mathrm{B}}B$. The entropy plot at $k_{{\rm B}}T/J_{1}=0.05$
displays the corresponding finite-temperature structure.}
\end{figure}

We adopt the single-cell basis $|S_{a,k},S_{b,k};s_{k}\rangle$, with
each spin variable $S_{a,k},S_{b,k},s_{k}=\pm\tfrac{1}{2}$. For a
chain of $N$ unit cells the Hamiltonian decomposes into local blocks,
and the ground-state configurations of these blocks determine the
zero-temperature phase diagram. Frustrated ground-state manifolds
may exhibit extensive degeneracy, leading to a finite residual entropy
$\mathcal{S}_{0}\equiv\lim_{T\to0}\mathcal{S}(T)$. At low but finite
temperatures, this residual entropy manifests itself as nearly constant
entropy plateaus. The low-temperature entropy map shown in Fig. \ref{fig:Ph-dgm1}
provides a direct visualization of this behavior: nonfrustrated regions
appear in cyan, with vanishing entropy, while frustrated sectors form
green and orange plateaus with $\mathcal{S}/k_{\mathrm{B}}\simeq\tfrac{1}{2}\ln2$
and $\mathcal{S}/k_{\mathrm{B}}\simeq\ln2$, respectively.

\subsection{Ground states}

For later comparison it is convenient to distinguish the sublattice
polarizations $\langle s_{k}\rangle$, $\langle S_{a,k}\rangle=\langle S_{b,k}\rangle\equiv\langle S_{k}\rangle$,
which varies between $-\tfrac{1}{2}$ and $\tfrac{1}{2}$, while the
total magnetization per cell becomes $M=\mu_{{\rm B}}(g_{0}\langle s_{k}\rangle+2g_{1}\langle S_{k}\rangle)$.

\subsubsection{Saturated phase (SP)}

At sufficiently strong magnetic field the system becomes saturated
phase (SP) or fully polarized
\begin{alignat*}{1}
\big|\mathrm{SP}\big\rangle= & \bigotimes_{k=1}^{N}\big|\tfrac{1}{2},\tfrac{1}{2};\tfrac{1}{2}\big\rangle_{k}.
\end{alignat*}
The energy per cell is 
\begin{alignat*}{1}
\varepsilon_{\mathrm{SP}} & =\frac{J_{2}}{4}+\frac{J_{0}+2J_{1}}{2}-\frac{2h_{1}+h_{0}}{2}.
\end{alignat*}
The local spin polarizations are $\langle s_{k}\rangle=\tfrac{1}{2}$
and $\langle S_{k}\rangle=\tfrac{1}{2}$, so the magnetization per
cell is $M=\mu_{{\rm B}}\tfrac{2g_{1}+g_{0}}{2}$. Since the ground
state is unique, the residual entropy vanishes, $\mathcal{S}_{\mathrm{SP}}=0$.
In Fig. \ref{fig:Ph-dgm1} the SP phase occupies the high-field plateaus,
where the entropy drops to zero.

\subsubsection{Partially frustrated phase}

The frustrated phase ($\mathrm{FR}_{1}$) has a period-two structure
in which one cell is fully polarized, while the next remains locally
frustrated. One representative configuration is
\begin{equation}
\big|\mathrm{FR}_{1}\big\rangle\!=\!\bigotimes_{k=1}^{N/2}\Big|\tfrac{1}{2},\tfrac{1}{2};\tfrac{1}{2}\!\Big\rangle_{2k-1}\Big|\eta_{2k},-\eta_{2k};-\tfrac{1}{2}\!\Big\rangle_{2k},
\end{equation}
with $\eta_{k}=\pm\tfrac{1}{2}$, where every second dimer carries
a free Ising orientation. Averaging over two consecutive cells gives
the energy per cell 
\begin{equation}
\varepsilon_{\mathrm{FR}_{1}}=\frac{J_{0}}{4}-\frac{J_{1}}{2}-\frac{h_{1}}{2}.
\end{equation}
The sublattice polarizations are $\langle S_{k}\rangle=\tfrac{1}{4}$
and $\langle s_{k}\rangle=0$. The corresponding magnetization per
cell is $M=\mu_{\mathrm{B}}(g_{0}\langle s_{k}\rangle+2g_{1}\langle S_{k}\rangle)=\mu_{\mathrm{B}}\frac{g_{1}}{2}$.
Because each frustrated cell carries one undetermined dimer orientation,
and assuming an even number of unit cells, the ground state has $2^{N/2}$configurations,
which yields the residual entropy $\mathcal{S}/k_{\mathrm{B}}=\tfrac{1}{2}\ln(2)$.
In Fig. \ref{fig:Ph-dgm1} this phase forms the intermediate-entropy
plateau (green region) with $\mathcal{S}/k_{\mathrm{B}}\simeq\tfrac{1}{2}\ln(2).$

\subsubsection{Fully frustrated phase}

The second frustrated ($\mathrm{FR}_{2}$) occurs when all nodal spins
are polarized upward, $s_{k}=\tfrac{1}{2}$, while the dimer spins
remain antiparallel. A convenient representative ground-state configuration
is  
\begin{equation}
\big|\mathrm{FR}_{2}\big\rangle=\bigotimes_{k=1}^{N}\Big|\eta_{k},-\eta_{k};\tfrac{1}{2}\Big\rangle_{k},\quad\eta_{k}=\pm\tfrac{1}{2}.
\end{equation}
Inserting this configuration into the Eq.\eqref{eq:Ham1} yields the
energy per cell
\begin{equation}
\varepsilon_{\mathrm{FR}_{2}}=-\frac{J_{2}}{4}-\frac{h_{0}}{2}.
\end{equation}

The dimer spin polarization vanishes, $\langle S_{k}\rangle=0$, while
the nodal sublattice is fully polarized, $\langle s_{k}\rangle=\tfrac{1}{2}$.
The corresponding magnetization per cell is therefore $M=\mu_{\mathrm{B}}\frac{g_{0}}{2}$.

Each dimer orientation $\eta_{k}=\pm\tfrac{1}{2}$ is independent,
the ground-state degeneracy is $2^{N}$, giving the residual entropy
$\mathcal{S}_{\mathrm{FR}_{2}}/k_{\mathrm{B}}=\ln(2)$. In Fig.\ref{fig:Ph-dgm1},
this phase corresponds to the high-entropy plateau (orange region),
where the entropy reaches $\mathcal{S}/k_{\mathrm{B}}\simeq\ln(2)$
at temperature $k_{{\rm B}}T/J_{1}=0.05$.

\subsubsection{Ferrimagnetic phase}

The ferrimagnetic phase (FI) is characterized by fully polarized dimer
spins and nodal spins aligned antiparallel to them. A representative
configuration is
\begin{equation}
|\mathrm{FI}\rangle=\bigotimes_{k=1}^{N}|\tfrac{1}{2},\tfrac{1}{2};-\tfrac{1}{2}\rangle_{k}.
\end{equation}
 The corresponding energy density is
\begin{equation}
\varepsilon_{\mathrm{FI}}=\frac{J_{2}}{4}-\frac{J_{0}+2J_{1}}{2}-h_{1}+\frac{h_{0}}{2}.
\end{equation}
 The sublattice spin polarizations are $\langle S_{k}\rangle=1/2$,
$\langle s_{k}\rangle=-1/2$. The corresponding physical magnetization
per cell is $M=\mu_{\mathrm{B}}\frac{2g_{1}-g_{0}}{2}$. As the configuration
is unique, the residual entropy vanishes, $\mathcal{S}_{\mathrm{AF}}=0$.
In Fig. \ref{fig:Ph-dgm1}, the FI phase appears as the ordered region
separating the AF and $\mathrm{FR}_{1}$ sectors.

\subsubsection{Antiferromagnetic phase}

Finally, the antiferromagnetic (AF) phase exhibits a period-two alternation
between fully polarized and fully reversed unit cells,
\begin{equation}
|\mathrm{AF}\rangle=\bigotimes_{k=1}^{N/2}|\tfrac{1}{2},\tfrac{1}{2};\tfrac{1}{2}\rangle_{2k-1}|-\tfrac{1}{2},-\tfrac{1}{2};-\tfrac{1}{2}\rangle_{2k}.
\end{equation}

with energy density given by
\begin{equation}
\varepsilon_{\mathrm{AF}}=\frac{J_{0}}{2}-J_{1}+\frac{J_{2}}{4}.
\end{equation}

Since the spin contributions from the two sublattices cancel exactly,
the nodal and dimer polarizations vanish, $\langle s_{k}\rangle=0$
and $\langle S_{k}\rangle=0$, while the total magnetization per cell
is therefore $M=0$. This configuration is unique up to global inversion,
therefore the residual entropy is zero, $\mathcal{S}_{\mathrm{AF}}/k_{\mathrm{B}}=0$.
In Fig. \ref{fig:Ph-dgm1}, the AF region appears immediately below
the FI region, forming the low-entropy sector adjacent to the frustrated
domains.

Taken together, the five phases SP, FI, AF, $\mathrm{FR}_{1}$, and
$\mathrm{FR}_{2}$ account for all ground-state regions observed in
Fig. \ref{fig:Ph-dgm1}. The frustrated sectors produce the characteristic
entropy plateaus, while the nondegenerate ordered regions is identified
as low-entropy (cyan regions). Their boundaries constitute the zero-temperature
skeleton underlying the sharp but continuous low-temperature crossovers
discussed in the next subsection.

\subsection{Phase boundary }

The analytic phase boundaries for the symmetric-field case $g_{0}=g_{1}$
follow from equating the corresponding ground-state energy densities
and reproduce exactly the straight dashed lines observed in Fig. \ref{fig:Ph-dgm1}.
Since $h_{0}=h_{1}=\mu_{\mathrm{B}}g_{0}B$, the phase diagram can
be expressed directly in the $(J_{2}/J_{1},\mu_{\mathrm{B}}g_{0}B/J_{1})$
plane.

Two boundaries do not depend on $J_{2}$ and therefore appear as horizontal
lines,
\begin{alignat*}{1}
\mathrm{FI}-\mathrm{AF} & :\ -h_{1}+\frac{h_{0}}{2}=J_{0},\quad\mathcal{S}/k_{{\rm B}}=0,\\
\mathrm{FI}-\mathrm{SP} & :\ h_{0}=J_{0}+2J_{1},\quad\mathcal{S}/k_{{\rm B}}=\ln(2),
\end{alignat*}
reflecting the fact that the competing phases have identical dependence
on the intra-dimer coupling $J_{2}$. The remaining boundaries have
nonzero slope. They are
\begin{alignat*}{1}
\mathrm{FR}_{1}-\mathrm{AF} & :\;h_{0}=\frac{-J_{0}+2J_{1}-J_{2}}{2},\quad\mathcal{S}/k_{{\rm B}}=\tfrac{1}{2}\ln(3),\\
\mathrm{FR}_{2}-\mathrm{SP} & :\;h_{1}=\frac{J_{0}}{2}+J_{1}+\frac{J_{2}}{2},\quad\mathcal{S}/k_{{\rm B}}=\ln(3),
\end{alignat*}
where in both cases the factor $\ln(3)$ arises because the competing
configurations at the phase boundary generate an additional local
degree of freedom not present in the bulk phases.

These lines are accompanied by a short slanted segment in the interval
$J_{2}\in[3J_{0}+2J_{1},\;2J_{1}-J_{0}]$, given by
\begin{alignat*}{1}
\mathrm{FR}_{1}-\mathrm{SP} & :h_{0}+h_{1}=\frac{J_{2}+6J_{1}+J_{0}}{4},\quad\mathcal{S}/k_{{\rm B}}=\ln(2).
\end{alignat*}

This segment corresponds precisely to the narrow oblique boundary
clearly visible between the $\mathrm{FI}-\mathrm{FR}_{1}$ and $\mathrm{FR}_{1}-\mathrm{FR}_{2}$
lines in Fig. \ref{fig:Ph-dgm1}. 

Finally, there are two peculiar boundaries,
\begin{alignat*}{1}
\mathrm{FI}-\mathrm{FR}_{1} & :\ h_{1}-h_{0}=\frac{J_{2}-3J_{0}-2J_{1}}{2},\\
\mathrm{FR}_{1}-\mathrm{FR}_{2} & :\ h_{1}-h_{0}=\frac{J_{2}+J_{0}-2J_{1}}{2},
\end{alignat*}
 and these do not exhibit any enhancement of the boundary residual
entropy. It is worth noting that both of these boundaries produce
clear pseudotransition signatures at low temperature. In Fig. \ref{fig:Ph-dgm1}
they manifest as sharp entropic ridges that separate the ordered and
frustrated regions in the $\mathrm{FI}-\mathrm{FR}_{1}$ case, and
similarly as a distinct ridge between the two frustrated states in
the $\mathrm{FR}_{1}-\mathrm{FR}_{2}$ case.

\subsection{Nearly block-diagonal form of the transfer matrix}

Since the transfer matrix $\mathbf{V}$ given in Eq. \eqref{eq:V}
is irreducible, the Non-existence of phase transition theorem\citep{cuesta}
guarantees the absence of true phase transitions. Nevertheless, $\mathbf{V}$
may become nearly block diagonal in certain parameter regimes. In
this case, the approach proposed in Ref. \citep{gen-conds} provides
a useful framework to analyze pseudocritical behavior from a general
perspective, without this tool it would become a cumbersome task find
the pseudocritical temperature. Following this approach, we reorganize
the transfer matrix into a nearly block-diagonal form in order to
identify the competing thermodynamic sectors responsible for the observed
pseudocritical temperatures.

\subsubsection{Pair-sector decomposition}

The transfer matrix $\mathbf{V}$ is written in the pair basis $\{|++\rangle,|+-\rangle,|-+\rangle,|--\rangle\}$,
where each state labels two consecutive nodal spins $(s_{k},s_{k+1})$.

This basis can be conveniently splits into two distinct sectors:

(i) Parallel sector $\{|++\rangle,\,|--\rangle\}$, corresponding
to locally uniform spin configurations.

(ii) Antiparallel sector $\{|+-\rangle,\,|-+\rangle\}$, corresponding
to locally alternating nodal spin configurations.

In terms of the ground-state patterns discussed above, the $\mathrm{FR}_{2}$
and FI manifolds are predominantly supported on parallel sector, whereas
the $\mathrm{FR}_{1}$ manifold is supported on antiparallel sector.
Consequently, the $\mathrm{FI}-\mathrm{FR}_{1}$ and $\mathrm{FR}_{1}-\mathrm{FR}_{2}$
pseudotransitions can be viewed as successive competitions between
these two pair sectors.

\subsubsection{Nearly block-diagonal structure of the transfer matrix}

Reordering the pair basis as $\{|++\rangle,|--\rangle\}\oplus\{|+-\rangle,|-+\rangle\}$,
the transfer matrix can be written in the block form

\begin{equation}
\mathbf{V}=\begin{pmatrix}\mathbf{A} & \mathbf{C}\\
\mathbf{D} & \mathbf{B}
\end{pmatrix},
\end{equation}
where the diagonal blocks are given by

\begin{alignat}{1}
\mathbf{A}=\begin{pmatrix}\mathfrak{r} & 0\\
0 & \mathfrak{t}
\end{pmatrix}, & \qquad\mathbf{B}=\begin{pmatrix}0 & \mathfrak{p}\\
\mathfrak{q} & 0
\end{pmatrix},
\end{alignat}
and the off diagonal blocks read
\begin{equation}
\mathbf{C}=\begin{pmatrix}\mathfrak{m} & 0\\
0 & \mathfrak{n}
\end{pmatrix},\qquad\mathbf{D}=\begin{pmatrix}0 & \mathfrak{n}\\
\mathfrak{m} & 0
\end{pmatrix}.
\end{equation}

The diagonal blocks $\mathbf{A}$ and $\mathbf{B}$ propagate configurations
within the parallel and antiparallel sectors, respectively, while
the off-block matrices $\mathbf{C}$ and $\mathbf{D}$ encode processes
that convert one sector into the other, which may be interpreted as
domain-wall-like events.

Although $\mathbf{V}$ is irreducible, the inter-sector couplings
$\mathbf{C}$ and $\mathbf{D}$ can become parametrically small in
the low-temperature regime. In this sense, the transfer matrix is
nearly block diagonal: the dominant statistical weight is carried
by one sector at a time, and pseudocritical behavior emerges when
the leading eigenvalues associated with different sectors become nearly
degenerate. This nearly reducible structure provides the spectral
mechanism underlying the sharp yet analytic crossovers discussed in
the main text.

The eigenvalues of the block $\mathbf{A}$ are trivially $\lambda=\{\mathfrak{r},\ \mathfrak{t}\},$
while the eigenvalues of the block $\mathbf{B}$ are $\lambda=\pm\sqrt{\mathfrak{p}\mathfrak{q}}$.

\subsubsection{Pseudocritical temperature}

According to the criterion discussed in Ref. \citep{gen-conds}, pseudocritical
behavior arises when the transfer matrix acquires a nearly block-diagonal
structure, such that the leading eigenvalues associated with competing
blocks become comparable. In the present case, this leads to two possible
pseudocritical conditions, depending on which eigenvalue of the block
$\mathbf{A}$ is dominant:
\begin{alignat}{1}
\mathfrak{t}=\sqrt{\mathfrak{p}\mathfrak{q}} & \quad\Rightarrow\;T_{p_{1}}:\quad\text{associted to}\quad\mathrm{FR}_{1}-\mathrm{FI},\label{eq:ex-tp1}\\
\mathfrak{r}=\sqrt{\mathfrak{p}\mathfrak{q}} & \quad\Rightarrow\;T_{p_{2}}:\quad\text{associted to}\quad\mathrm{FR}_{2}-\mathrm{FI}.\label{eq:ex-pt2}
\end{alignat}
These relations provide a transparent interpretation of the two pseudocritical
temperature scales observed in the thermodynamic analysis. It is worth
emphasizing that they follow directly from the criterion of Ref. \citep{gen-conds}.
Although simple in form, the resulting conditions involve transcendental
equations.

Alternatively, the pseudocritical temperature $T_{p}$ can also be
determined along any fixed parameter cut in the low-temperature regime.
For instance at fixed $J_{0}$, $J_{1}$, $J_{2}$, while varying
$B$, follows from the crossing of the low temperature free energy
branches of the two competing phases. Approximating each free energy
by $f_{A}(T)\simeq\varepsilon_{A}-k_{\mathrm{B}}Ts_{A}$, the condition
$f_{A}(T_{\mathrm{p}})=f_{B}(T_{\mathrm{p}})$ yields the standard
estimate of pseudocritical temperature
\begin{equation}
k_{\mathrm{B}}T_{\mathrm{p}}\approx\frac{\varepsilon_{B}-\varepsilon_{A}}{\mathcal{S}_{B}-\mathcal{S}_{A}}.
\end{equation}
Applying this expression to the competition between $\mathrm{FI}$
and $\mathrm{FR}_{1}$ gives 
\begin{equation}
T_{p_{1}}\approx\frac{3J_{0}+2J_{1}-J_{2}+2(h_{1}-h_{0})}{2k_{\mathrm{B}}\ln(2)},\label{eq:Tp1}
\end{equation}
which accurately reproduces the low-temperature behavior of $T_{p_{1}}$
obtained in Eq. \eqref{eq:ex-tp1}. In a similarly manner, the pseudocritical
temperature associated with the $\mathrm{FR}_{1}$ and $\mathrm{FR}_{2}$
competition is estimated as 
\begin{equation}
T_{p_{2}}\approx\frac{2J_{1}-J_{0}-J_{2}+2(h_{1}-h_{0})}{2k_{{\rm B}}\ln(2)}.\label{eq:Tp2}
\end{equation}
 in good agreement with Eq. \eqref{eq:ex-pt2} in the low temperature
region.

Both expressions vanish linearly as the parameters approach their
respective zero-temperature boundaries. If $g_{0}=g_{1}$, $T_{p_{1}}$
and $T_{p_{2}}$ are independent of $B$, and appears as vertical
lines in Fig.\ref{fig:Ph-dgm1}. 

\begin{figure}
\includegraphics[scale=0.52]{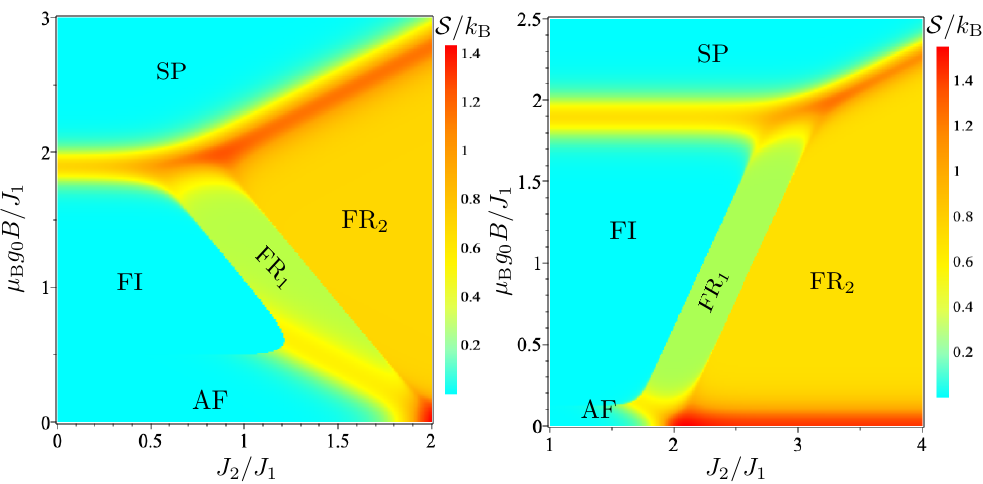}\caption{\label{fig:Ph-dgm2}Low-temperature phase diagram of the CTDC for
$g_{0}\protect\ne g_{1}$ under a uniform external field $B$, assuming
fixe $J_{0}/J_{1}=-0.1$. (left) Dimer spins have weaker Zeeman response,
$g_{1}/g_{0}=0.7$. (right) Dimer spins have stronger Zeeman response,
$g_{1}/g_{0}=1.3$. The entropy plot at $k_{{\rm B}}T/J_{1}=0.05$
displays the corresponding finite-temperature structure.}
\end{figure}

In Fig. \ref{fig:Ph-dgm2} is shown, how the low-temperature entropy
$S(T)$ landscape changes in the $(J_{2}/J_{1},\,\mu_{{\rm B}}g_{0}B/J_{1})$
plane, when the two sublattices have $g_{0}\ne g_{1}$. In panel (left),
where $g_{1}/g_{0}>1$, the frustrated boundaries that appear as vertical
lines in the symmetric case tilt into straight lines with a positive
slope. In panel (right), where $g_{1}/g_{0}<1$, the same boundaries
acquire a negative slope, producing a opposite deformation of the
diagram.

Although the geometric orientation of the boundaries changes, the
entropy plateaus remain unchanged: the ordered phases occupy the low-entropy
cyan regions, while the frustrated $\mathrm{FR}_{1}$ and $\mathrm{FR}_{2}$
exhibit green and orange regions characteristic values $\mathcal{S}/k_{\mathrm{B}}=\tfrac{1}{2}\ln(2)$
and $\ln(2)$, respectively. The $\mathrm{FR}_{1}-\mathrm{SP}$ oblique
boundary also persists as a distinct entropic ridge in both panels,
although shifted and tilted by the asymmetric Zeeman response.

Importantly, the deformation of these boundaries does not suppress
the thermal anomalies associated with the frustrated manifold. The
tilted ridges that separate FI from $\mathrm{FR}_{1}$, and $\mathrm{FR}_{1}$
from $\mathrm{FR}_{2}$ continue to generate clear pseudotransition
signatures at low temperature, reflected in the sharp entropic gradients
that survive in both panels. 

\section{Thermodynamics and pseudotransition}

Here we examine the thermodynamic and magnetic properties in the vicinity
of the anomalous crossover.

\subsection{Entropy and specific heat}

\begin{figure}
\includegraphics[scale=0.5]{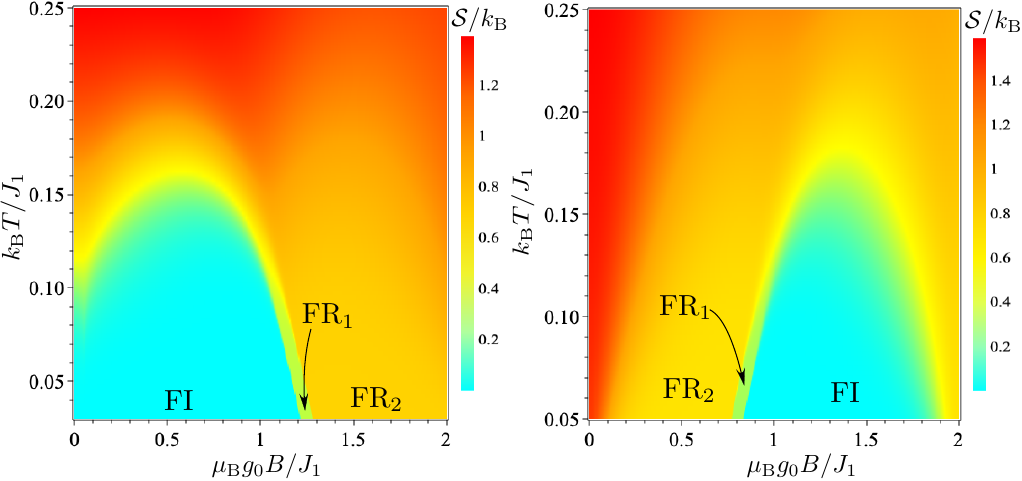}\caption{\label{fig:Entropy-h}Density plot of the entropy in the $(\mu_{{\rm B}}g_{0}B/J_{1},\,k_{{\rm B}}T/J_{1})$
plane, for a weak ferromagnetic coupling $J_{0}$. (left) $J_{0}/J_{1}=-0.01$,
$J_{2}/J_{1}=1.2$ and $g_{1}/g_{0}=0.7$; (right) Similarly for $J_{0}/J_{1}=-0.01$,
$J_{2}/J_{1}=2.4$ and $g_{1}/g_{0}=1.3$.}
\end{figure}

In Fig. \ref{fig:Entropy-h} is reported the low-temperature entropy
landscape in the $(\mu_{{\rm B}}g_{0}B/J_{1},\,k_{{\rm B}}T/J_{1})$
plane for $g_{1}\neq g_{0}$. Panel (left), obtained for $J_{0}/J_{1}=-0.01$,
$J_{2}/J_{1}=1.2$ and $g_{1}/g_{0}=0.7$, shows three distinct low-temperature
domains that reproduce the structure of the ground-state phase diagram.
A narrow green plateau associated with the $\mathrm{FR}_{1}$ phase
appears between the FI and $\mathrm{FR}_{2}$ sectors, and its boundary
acquires a negative slope due to the asymmetric Zeeman response. In
this region the entropy ridges separating the ordered and frustrated
sectors mark the emergence of two pseudotransition scales, one related
to the competition between FI and $\mathrm{FR}_{1}$ and another associated
with the competition between $\mathrm{FR}_{1}$ and $\mathrm{FR}_{2}$.
These features become progressively smoother as the temperature increases.
Panel (right), corresponding to $J_{0}/J_{1}=-0.01$, $J_{2}/J_{1}=2.4$
and $g_{1}/g_{0}=1.3$, shows the opposite situation where the frustrated
boundaries tilt with positive slope, reflecting the opposite imbalance
in the magnetic response. The pseudotransition ridges remain visible,
indicating that the entropic signatures of the frustrated manifold
are robust against variations in the ratio $g_{1}/g_{0}$.

\begin{figure}
\includegraphics[scale=0.52]{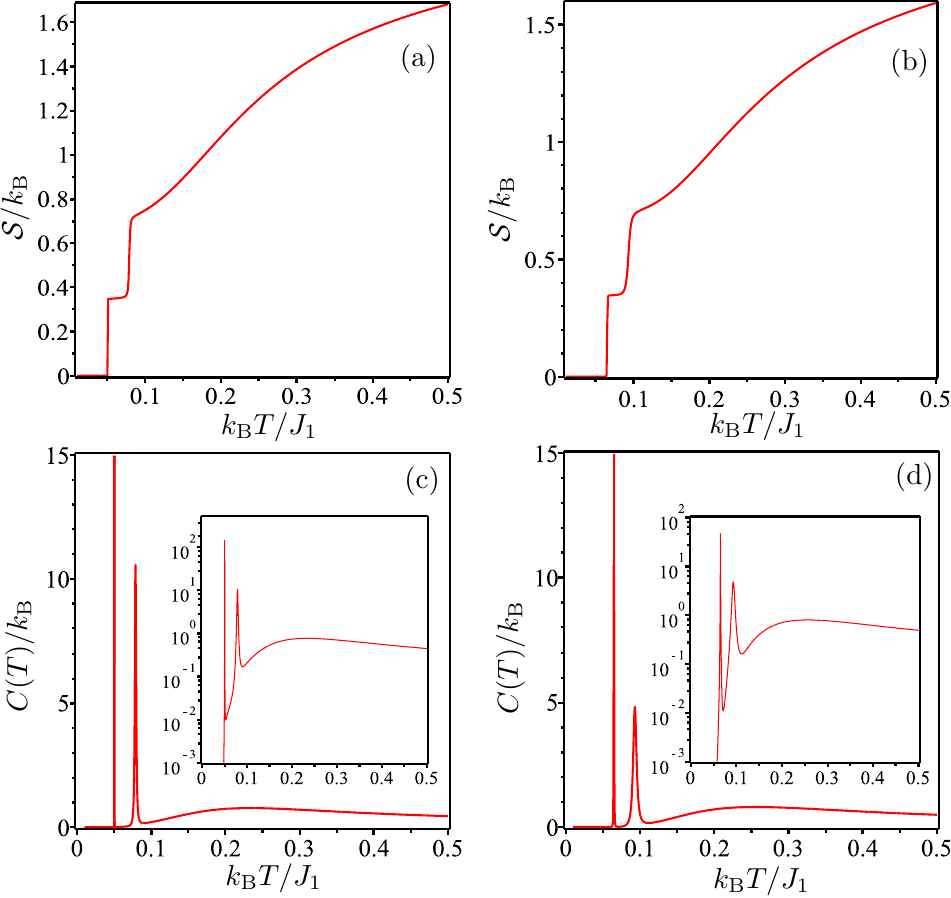}\caption{\label{fig:S-and-C}(a) Entropy as functions of temperature for $J_{2}/J_{1}=1.24$,
$J_{0}/J_{1}=-0.01$, $\mu_{{\rm B}}g_{0}B/J_{1}=1.1$, and $g_{1}/g_{0}=0.7$.(b)
Entropy as functions of temperature for $J_{2}/J_{1}=2.48$, $J_{0}/J_{1}=-0.01$,
$\mu_{{\rm B}}g_{0}B/J_{1}=1$, and $g_{1}/g_{0}=1.3$. (c) Specific
heat for the parameters in panel (a), with an inset showing $C(T)$
on a logarithmic temperature scale.(d) Specific heat for the parameters
in panel (b), with an inset showing $C(T)$ on a logarithmic temperature
scale.}

\end{figure}

Figure \ref{fig:S-and-C} shows the temperature dependence of the
entropy and specific heat for two representative parameter sets reported
in the caption of Fig.\ref{fig:S-and-C}, illustrating the emergence
of two distinct pseudocritical temperatures. These scales are associated
with the $\mathrm{FI}$ to $\mathrm{FR}_{1}$ crossover, with pseudocritical
temperature given by Eq.\eqref{eq:ex-tp1} at $k_{{\rm B}}T_{p_{1}}/J_{1}\approx0.0504764$,
and with the $\mathrm{FR}_{1}$ to $\mathrm{FR}_{2}$ crossover, with
pseudocritical temperature given by Eq. \eqref{eq:ex-pt2} at $k_{{\rm B}}T_{p_{2}}/J_{1}\approx0.0788948$.
Panels (a) and (b) display the entropy curves, which show two smooth
but clearly separated increases as temperature rises. The lower increase
at $T_{p_{1}}$ corresponds to the partial activation of the $\mathrm{FR}_{1}$
manifold, whereas the higher increase at $T_{p_{2}}$ signals access
to the fully frustrated $\mathrm{FR}_{2}$ sector.

Panels (c) and (d) present the corresponding specific heat $C(T)$,
which develops two continuous maxima at $T_{p_{1}}$ and $T_{p_{2}}$.
The logarithmic insets make the locations of these maxima particularly
visible. The first maximum marks the crossover between the ordered
FI region and the partially frustrated $\mathrm{FR}_{1}$ configurations,
while the second reflects the competition between $\mathrm{FR}_{1}$
and $\mathrm{FR}_{2}$. We can observe that the peak at $T_{p_{1}}$
is higher and sharper than the second peak at $T_{p_{2}}$. This behavior
is consistent with the pseudocritical features reported in the literature\citep{sergio18,Rojas2019,katarina,verissimo}.

The clear separation between these double thermal anomalies shows
that the model supports two well defined pseudocritical temperature
scales, each associated with a distinct degeneracy change in the ground-state
structure.

\subsection{Magnetic properties}

\begin{figure}
\includegraphics[scale=0.52]{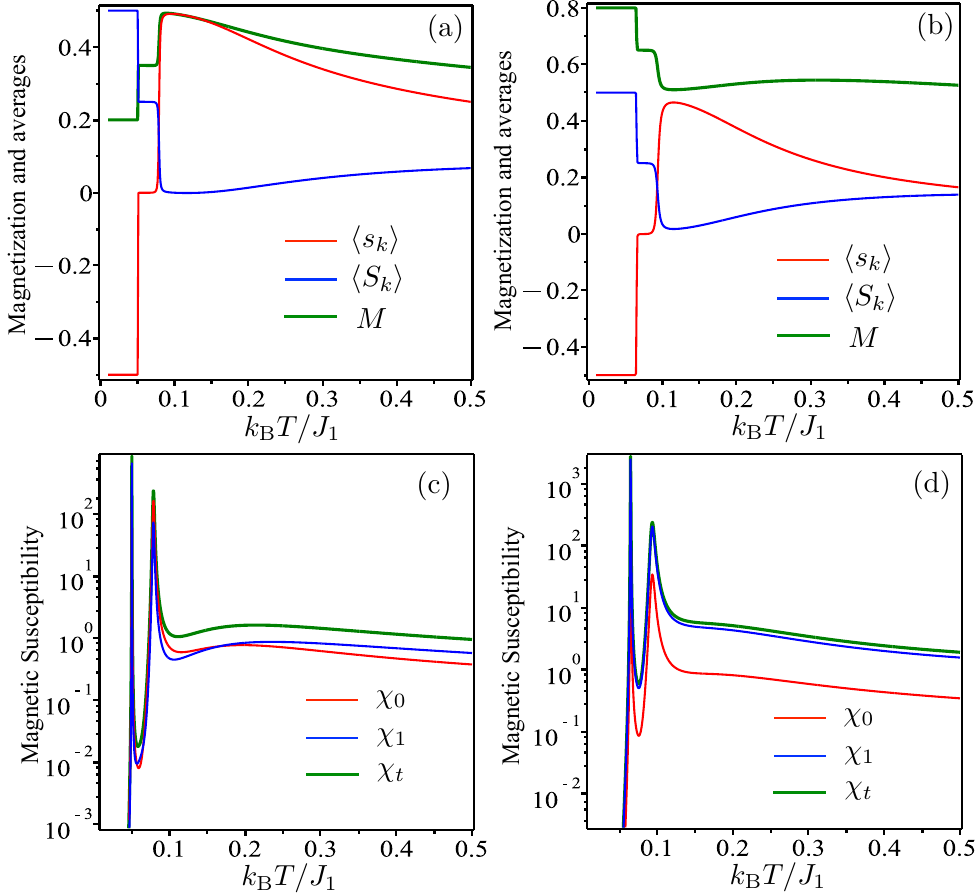}\caption{\label{fig:MXT}Magnetization ($M)$ and averages $\langle s_{k}\rangle$
and $\langle S_{k}\rangle$, as functions of temperature. (a) For
$J_{2}/J_{1}=1.24$, $J_{0}/J_{1}=-0.01$, $\mu_{{\rm B}}g_{0}B/J_{1}=1.1$,
and $g_{1}/g_{0}=0.7$. (b) For$J_{2}/J_{1}=2.48$, $J_{0}/J_{1}=-0.01$,
$\mu_{{\rm B}}g_{0}B/J_{1}=1$, and $g_{1}/g_{0}=1.3$. (c) Magnetic
susceptibility for the parameters in panel (a). (d) Magnetic susceptibility
for the parameters in panel (b).}
\end{figure}

Figure \ref{fig:MXT} shows the temperature dependence of the magnetization,
the sublattice averages, and the magnetic susceptibility for the same
parameter sets used in Fig. \ref{fig:S-and-C}. 

Panels (a) and (b) display $\langle s_{k}\rangle$, $\langle S_{k}\rangle$
and $M$ as functions of temperature. All three curves show two smooth
but clearly separated steps. The first step, located at the pseudocritical
temperature $T_{p_{1}}$, reflects the partial activation of the $\mathrm{FR}_{1}$
configurations, which introduce local frustration while keeping the
nodal subsystem partially aligned. The second step, located at $T_{p_{2}}$,
marks the activation of the fully frustrated $\mathrm{FR}_{2}$ manifold,
where each dimer contributes an additional independent degree of freedom.

The dimer and nodal magnetic susceptibilities are defined through
the derivative of the free energy with respect to the external field
$B$, namely $\chi_{0}(T)=\mu_{\mathrm{B}}g_{0}\,\frac{\partial\langle s_{k}\rangle}{\partial B}$,
$\chi_{1}(T)=2\mu_{\mathrm{B}}g_{1}\,\frac{\partial\langle S_{k}\rangle}{\partial B}$,
whereas the total magnetic susceptibility is $\chi_{t}(T)=\chi_{0}(T)+\chi_{1}(T)$.

Panels (c) and (d) of Fig. \ref{fig:MXT} show $\chi_{0}$, $\chi_{1}$,
and $\chi_{t}$ on a logarithmic scale. All three susceptibilities
develop two pronounced peaks at $T_{p_{1}}$ and $T_{p_{2}}$. These
peaks coincide with the pseudocritical temperatures identified from
the entropy and specific-heat analysis. Although the peaks appear
sharp, they remain finite and represent the typical thermal signatures
of pseudotransitions in frustrated one-dimensional systems. Again,
the anomalies occur at the same temperatures identified from the entropy
and specific heat, and the peak at $T_{p_{1}}$ is higher and sharper
than the one at $T_{p_{2}}$. This confirms the consistency of the
dual pseudotransition behavior in the CTDC model.

\subsection{Eigenvalues properties and correlation length}

Figure \pageref{fig:lmbds-xi}(a) shows the temperature dependence
of the relative eigenvalues $|\lambda_{j}|/\lambda_{1}$ for the parameter
set $J_{2}/J_{1}=1.24$, $J_{0}/J_{1}=-0.01$, $\mu_{\mathrm{B}}g_{0}B/J_{1}=1.1$,
and $g_{1}/g_{0}=0.7$. Although the eigenvalues satisfy the ordering
$\lambda_{1}>\lambda_{2}>\lambda_{3}>\lambda_{4}$, the subleading
eigenvalue in modulus is not $\lambda_{2}$ but rather $|\lambda_{4}|$
(see appendix \ref{sec:appx-A}). This behavior is illustrated by
the red curve $|\lambda_{4}|/\lambda_{1}$, while the green and blue
curves correspond to $\lambda_{2}/\lambda_{1}$ and $\lambda_{3}/\lambda_{1}$,
respectively. The ratio $|\lambda_{4}|/\lambda_{1}$ exhibits a broad
plateau between the two pseudocritical temperatures, indicating that
the same low-energy sector remains dominant throughout this interval.
In contrast, the ratio $\lambda_{2}/\lambda_{1}$ displays two pronounced
extrema located precisely at the pseudocritical temperatures $T_{p_{1}}$
and $T_{p_{2}}$, reflecting the successive competition between distinct
thermodynamic sectors.

For the same parameters, Fig. \pageref{fig:lmbds-xi}(b) shows the
quantities $\xi_{j}=\bigl[\ln(\lambda_{1}/|\lambda_{j}|)\bigr]^{-1}$
as functions of temperature on a logarithmic scale. Since $|\lambda_{4}|$
is the largest subleading eigenvalue in modulus, the spectral correlation
length is given by $\xi=\xi_{4}=\bigl[\ln(\lambda_{1}/|\lambda_{4}|)\bigr]^{-1}$.
However, this quantity does not exhibit signatures of the pseudocritical
behavior. By contrast, the quantity $\xi_{2}=\bigl[\ln(\lambda_{1}/\lambda_{2})\bigr]^{-1}$,
although not a correlation length in the strict spectral sense, clearly
signals both pseudocritical temperatures $T_{p_{1}}$ and $T_{p_{2}}$.
In general, the spectral correlation length is determined by

\begin{equation}
\xi=\left[\ln\left(\frac{\lambda_{1}}{\max\{|\lambda_{2}|,|\lambda_{3}|,|\lambda_{4}|\}}\right)\right]^{-1},
\end{equation}
 whereas pseudocritical behavior is associated with near-degeneracies
between the leading eigenvalues that govern the dominant thermodynamic
sectors.

\begin{figure}

\includegraphics[scale=0.5]{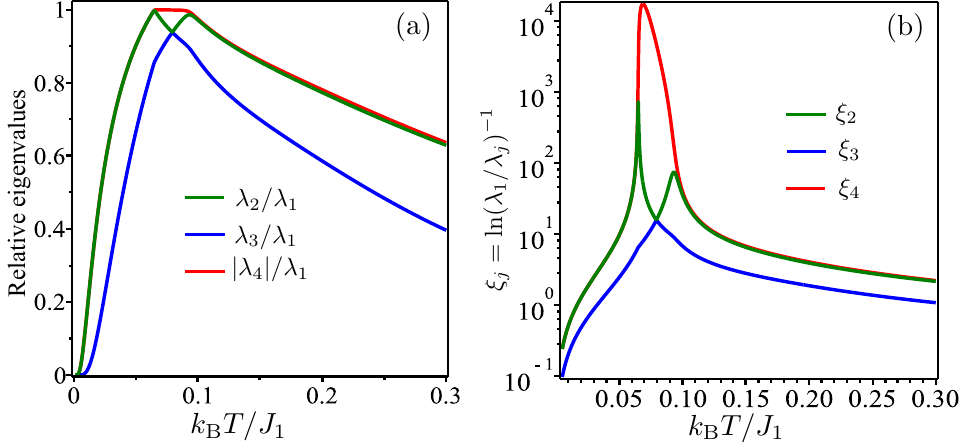}\caption{\label{fig:lmbds-xi}(a) Temperature dependence of the relative eigenvalues
$|\lambda_{j}|/\lambda_{1}$ for the fixed parameters $J_{2}/J_{1}=2.48$,
$J_{0}/J_{1}=-0.01$, $\mu_{{\rm B}}g_{0}B/J_{1}=1$, and $g_{1}/g_{0}=1.3$.
(b) Temperature dependence of the quantities $\xi_{j}=\ln(\lambda_{1}/\lambda_{j})^{-1}$
for the same set of parameters as in panel (a).}
\end{figure}

Although universality and quasicritical exponents have been identified
for a broad class of one-dimensional decorated models \citep{Rojas2019},
a quantitative scaling analysis of correlation lengths is not pursued
here. In contrast, the quantity $\xi_{2}=[\ln(\lambda_{1}/\lambda_{2})]^{-1}$
exhibits anomalous critical behavior, but a detailed analysis of its
associated exponents lies beyond the scope of the present work. 

The emergence of two pseudocritical temperatures in the present model
is closely related to the coupling of two diamond units within the
same chain. However, increasing the number of diamond units or coupling
additional chains does not, in general, guarantee the appearance of
further pseudocritical temperatures. In the coupled twin-diamond chain,
the internal structure of the unit cell supports two frustrated manifolds,
$\mathrm{FR_{1}}$ and $\mathrm{FR}_{2}$, with different extensive
degeneracies, leading to successive entropy-driven free-energy crossings
as temperature increases and hence to two well-separated pseudotransitions.
Additional pseudotransitions may arise only when the enlarged unit
cell stabilizes new competing low-energy sectors with distinct residual
entropies that survive the coupling, consistently with the criterion
discussed in Ref. \citep{bjp20}.

\section{Conclusion}

In this work we have presented an exact analysis of the coupled twin-diamond
chain (CTDC), a decorated one-dimensional Ising system inspired by
the magnetic structure of $\mathrm{Cu}_{2}(\mathrm{TeO}_{3})_{2}\mathrm{Br}_{2}$.
The model contains two inequivalent sublattices, allowing for distinct
local fields, and includes three exchange couplings that control the
interplay between nodal spins and internal dimers. Using the star-triangle
mapping\citep{dec-trns,ddec-trns,jozef-dctrns} and the resulting
$4\times4$ transfer matrix, we obtained analytic expressions for
the eigenvalues governing the full thermodynamics of the system.

The zero-temperature analysis reveals five distinct ground-state sectors:
the saturated phase (SP), the ferrimagnetic phase (FI), the antiferromagnetic
phase (AF), and two frustrated phases, $\mathrm{FR}_{1}$ and $\mathrm{FR}_{2}$.
Each phase is characterized by a simple product state, and their energy
densities, sublattice polarizations, and residual entropies were derived
explicitly. The two frustrated phases carry extensive degeneracies,
yielding residual entropies $\mathcal{S}/k_{\mathrm{B}}=\tfrac{1}{2}\ln(2)$
for $\mathrm{FR}_{1}$ and $\mathcal{S}/k_{\mathrm{B}}=\ln(2)$ for
$\mathrm{FR}_{2}$. These values are directly reflected in the low-temperature
entropy map, where $\mathrm{FR}_{1}$ and $\mathrm{FR}_{2}$ induce
the intermediate- and high-entropy plateaus regions, respectively.

The complete phase diagram in the ($J_{2}/J_{1},\mu_{{\rm B}}g_{0}B/J_{1}$)
plane shows that the frustrated sectors occupy extended regions between
the ordered phases. Because the boundaries between these sectors involve
ground-state level crossings accompanied by changes in degeneracy,
the associated finite-temperature behavior displays sharp but continuous
features. In particular, the entropy and magnetization exhibit steplike
but nondiscontinuous variations, while the specific heat and susceptibility
develop pronounced finite peaks-typical signatures of pseudotransition
behavior previously reported for other decorated Ising chains.

Altogether, the CTDC provides another exactly solvable example in
which competing local configurations and internal degeneracies generate
well-defined frustrated regions and clear thermodynamic precursors
of pseudotransitions at low temperature. The analytic structure offers
a clear framework for understanding how decoration, sublattice asymmetry,
and frustration generate dual pseudocritical scales in the CTDC model.

\appendix

\section{Pairwise structure of the eigenvalues\label{sec:appx-A}}

Here we clarify the algebraic structure of the four eigenvalues of
the transfer matrix and show that two eigenvalues associated with
the Perron branch, namely $\lambda_{1}$ and $\lambda_{2}$ given
in Eqs. \eqref{eq:L1} and \eqref{eq:L2}, are necessarily real, while
the remaining eigenvalues may be either real or form a complex-conjugate
pair.

\subsection{Pairwise algebraic structure of the spectrum}

The exact diagonalization yields the four eigenvalues in the form
\begin{alignat}{1}
\lambda_{1} & =c+\frac{R_{1}+\sqrt{R_{-}}}{2},\qquad\lambda_{2}=c+\frac{R_{1}-\sqrt{R_{-}}}{2},\label{eq:app-lmbd12}\\
\lambda_{3} & =c-\frac{R_{1}-\sqrt{R_{+}}}{2},\qquad\lambda_{4}=c-\frac{R_{1}+\sqrt{R_{+}}}{2},\label{eq:app-lmbd34}
\end{alignat}
with 
\begin{equation}
c=\frac{\mathfrak{r}+\mathfrak{t}}{4},\qquad R_{1}=\sqrt{2y-a_{2}},
\end{equation}
and 
\begin{equation}
R_{\mp}=-2y-a_{2}\mp\frac{2a_{1}}{R_{1}}.
\end{equation}
Here all coefficients are real, and the auxiliary variable $y$ is
chosen as a real root of the resolvent cubic.

Equation \eqref{eq:app-lmbd12} shows that the spectrum naturally
decomposes into two algebraic pairs: $(\lambda_{1},\lambda_{2})$,
controlled by the radicand $R_{-}$, and $(\lambda_{3},\lambda_{4})$,
controlled by $R_{+}$. Any complex eigenvalues can therefore arise
only within one of these pairs.

\subsection{Reality of $\lambda_{1}$ and $\lambda_{2}$}

The transfer matrix has strictly positive entries and is primitive.
By the Perron-Frobenius theorem \citep{cuesta}, the dominant eigenvalue
$\lambda_{1}$ is real, strictly positive, and nondegenerate, and
is given by the first expression in Eq. \eqref{eq:app-lmbd12}. If
$R_{-}<0$, then $\sqrt{R_{-}}$ would be purely imaginary, causing
$\lambda_{1}$ to acquire a nonzero imaginary part, which is impossible.
Therefore $R_{-}\geqslant0$, and both $\lambda_{1}$ and $\lambda_{2}$
are real on the thermodynamic branch.

\subsection{Remaining eigenvalues and possible complex conjugate pair}

No analogous restriction applies to the second radicand $R_{+}$.
If $R_{+}<0$, then $\sqrt{R_{+}}$ is purely imaginary, and Eq. \eqref{eq:app-lmbd12}
immediately yields $\lambda_{4}=\lambda_{3}^{*}$, so that $\lambda_{3}$
and $\lambda_{4}$ form a complex-conjugate pair. If $R_{+}\geqslant0$,
all four eigenvalues are real.

Thus, on the thermodynamic branch, the spectrum always contains at
least two real eigenvalues, $\lambda_{1}$ and $\lambda_{2}$, while
the remaining pair may be real or complex conjugate, depending on
the Hamiltonian parameters.

\subsection{Magnitude of eigenvalues}

Although the eigenvalues $\lambda_{j}$ are labeled according to their
algebraic expressions, their ordering on the real axis is parameter
dependent. Even when all eigenvalues are real, the second-largest
eigenvalue in magnitude need not coincide with $\lambda_{2}$; depending
on parameters, it may instead correspond to $\lambda_{3}$ or $\lambda_{4}$.
Consequently, subleading behavior is governed by the eigenvalue with
the largest modulus among $\lambda_{j\neq1}$, rather than by a fixed
algebraic branch.
\begin{acknowledgments}
This work was partially supported by the Brazilian agencies CNPq and
FAPEMIG.
\end{acknowledgments}

\end{document}